\documentclass[useAMS,usenatbib]{mn2e}
\usepackage{aas_macros,mathptmx}
\usepackage{psfig}

\def\lcdm{$\Lambda$CDM }

\title{Power Spectrum Normalisation and the Non-Gaussian Halo Model}
\author[Amara \& Refregier]{Adam Amara$^{1}$ \& Alexandre Refregier$^{2}$\\
$^{1}$Institute of Astronomy, Madingley Road, Cambridge CB3 OHA, UK;
  aa@ast.cam.ac.uk\\
$^{2}$Service d'Astrophysique, CEA Saclay, Gif sur Yvette, 91191, France;
refregier@cea.fr}

\date{Accepted ---. Received ---; in original form ---.}

\pagerange{\pageref{firstpage}--\pageref{lastpage}}
\pubyear{2001}
\begin{document}
\maketitle
\label{firstpage}

\begin{abstract}
The normalisation of the matter power spectrum, $\sigma_{8}$, is an
essential ingredient to predict the phenomenology of the low redshift
universe. It has been measured using several methods, such as X-ray
cluster counts, weak lensing and the cosmic microwave background,
which have yielded values ranging from 0.7 to 1.0. While these
differences could be due to systematic effects, they could also be due
to physics beyond the standard \lcdm model. An obvious possibility is
the presence of non-Gaussian initial fluctuations in the density
field. To study the impact of non-Gaussianity on each of these
methods, we use a generalised halo model to compute cluster counts and
the non-linear power spectrum for non-Gaussian models. Assuming scale
invariance, the upper-limits on non-Gaussianity from the WMAP CMB
experiment correspond to roughly a 4\% shift in $\sigma_8$
 as
measured from cluster counts and about 2\% shift through weak lensing.
This is not enough to account for the current internal and mutual
discrepancies between the different methods, unless non-Gaussianity is
strongly scale dependent. A comparison between future X-ray surveys
with a two fold improvement in cluster mass calibration and future
cosmic shear surveys with 400 deg$^{2}$ will be required to constrain
non-Gaussianity on small scales with a precision matching that of the
current CMB constraints on larger scales. Our results argue for the
presence of systematics in the current cluster and cosmic shear
surveys, or to non-standard physics other than non-Gaussianity.
\end{abstract}

\begin{keywords}
gravitation -- cosmology: dark matter -- theory -- 
large-scale structure of Universe
\end{keywords}

\section{Introduction}
\label{intro}
In the inflation-driven \lcdm paradigm, the initial fluctuations are
assumed to obey Gaussian statistics. Current observations of the
cosmic microwave background (CMB) of galaxy clustering and of galaxy
clusters are consistent with this hypothesis
\citep[eg.][]{2001MNRAS.325..412V,2003PhRvD..68b1302G}. However,
models with non-Gaussian initial conditions have been proposed and are
not yet ruled out by observations \citep[see][ and reference
therein]{2003astro.ph..6293A}.  These models include ones based on
inflation theories with non-linearities of a single scalar field
\citep{2000PhRvD..61h3518M,2002PhRvD..66h3502G} or with multi-fields
\citep{1999ApJ...510..523P,1997PhRvL..79...14A,1997PhRvD..56..535L,2002PhRvD..66j3506B},
or on cosmological defects
\citep{1990PhRvL..64.2736T,1998ApJ...507L.101A}.

In this paper, we study how non-Gaussian initial conditions can affect
the determination of $\sigma_{8}$, the normalisation of the matter
power spectrum on 8 $h^{-1}$ Mpc scales. This normalisation is an
essential ingredient to predict the phonemenology of the low redshift
universe and has been measured using several methods. The recent CMB 
measurements with WMAP
\citep{2003ApJS..148..175S} yield a constraint on the normalisation of
$\sigma_8=0.9\pm0.1$ (68\%CL). The abundance of galaxy clusters
depends strongly on $\sigma_{8}$ and yields values in the range of
$\sigma_8=0.7$-0.8 \citep[eg.][]{2003MNRAS.342..163P} for $\Omega_m
\approx 0.3$. Weak lensing by large-scale structure, or 'cosmic
shear', provides a direct measurement of mass fluctuations and has
recently been detected and measured \citep[see][~for recent
reviews]{2003astro.ph..7212R,2003astro.ph..5089V}. The different
cosmic shear surveys yield values of $\sigma_8$ in the range of
$0.7-0.9$ for the same value of $\Omega_m$. The marginal discrepancies
within and between these different techniques may be due to residual
systematics, such as uncertainties in the mass-temperature of X-ray
clusters, or the calibration of the shear in cosmic shear
surveys. They could also be due to physics beyond the standard \lcdm
model, which has different effects on the various methods and
surveys.  In particular, non-Gaussianity appears as a prime candidate
to explain discrepancies between the different methods. Cluster
abundance and cosmic shear indeed both probe non-linear structures at
low redshifts, but are sensitive to matter fluctuations on different
scales. Non-Gaussianity may thus affect these two methods differently,
while leaving the CMB power spectrum essentially
unaffected. Reciprocally, it is interesting to establish whether the
comparison between the different methods with present and future
surveys may constrain non-Gaussianity.

To study the impact of non-Gaussianity on measurements of
$\sigma_8$,
 we present a generalisation of the halo model
\citep{2000ApJ...543..503M,2000MNRAS.318..203S}.  Building on the
work
 of \cite{2000MNRAS.311..781R} for the mass function,
\cite{2000ApJ...543..503M} and \cite{1999MNRAS.310.1111K} for the
halo
 bias, and \cite{1996ApJ...462..563N} and
\cite{2003astro.ph..6293A} for the halo profiles, we compute the
non-linear power spectrum for arbitrary non-Gaussian models. This
allows us to compute the cluster temperature function and the cosmic
shear statistics for these models. We then study whether a
discrepancy in the determination of $\sigma_8$ from the different
observational methods can be explained by non-Gaussianity. We also
discuss how future surveys can be compared to set constraints on
non-Gaussianity. The paper is organised as follows. In
\S\ref{nongmods}, we present the different non-Gaussian models and
discuss how they are constrainted with current observations. In
\S\ref{halo_model}, we describe the non-Gaussian halo model. In
\S\ref{observations}, we study the effect of non-Gaussianity on the
determination of $\sigma_8$ with the different methods. Our
conclusions are summarised in \S\ref{conclusion}.
\section{Non-Gaussian Models}
\label{nongmods}

The statistics of the post-recombination linear density field $\rho$
are usually described by first forming the density contrast $\delta =
(\rho-\overline{\rho})/\overline{\rho}$ with respect to the average
matter density $\overline{\rho}=\langle \rho \rangle$ in the universe.
We consider the probability distribution function (PDF)
$p(\delta,M)$ of the density contrast $\delta$ smoothed with a
spherical top-hat of radius $R$ corresponding to an enclosed mass
$M=4\pi\overline{\rho} R^{3}/3$. It is defined so that
\begin{equation}
\label{eq:pdf_cond1}
\int d\delta~p(\delta,M) = 1,
~~\int d\delta~p(\delta,M)\delta = \langle \delta \rangle = 0.
\end{equation}
The variance of the distribution is given by
\begin{equation}
\label{eq:pdf_cond2}
\int d\delta~p(\delta,M)\delta^2 = \langle \delta^2 \rangle =
\sigma^{2}(M)
\end{equation}
and can be computed from the linear matter power spectrum $P_{\rm
lin}(k)$ using
\begin{equation}\label{eq:spec}
\sigma^{2}(M)=\int \frac{d^3k}{(2\pi)^{3}}~P_{\rm lin}(k)W^{2}(k,M),
\end{equation}
where $W(k,M)=3(\sin kR/kR-\cos kR)/(kR)^2$ is the top-hat window
function. The linear power spectrum is calculated using the transfer
function from \cite{1984ApJ...285L..45B}.

To quantify the level of non-Gaussianity of the distribution,
\cite{2000ApJ...532....1R} have considered the excess probability
$\alpha$ in the $3\sigma$ tail over that of a Gaussian PDF, i.e.
\begin{equation}
\label{eq:alpha}
\alpha(M)=\frac{\int_{3\sigma(M)}^{\infty}d\delta~p(\delta,M)}
  {\int_{3\sigma(M)}^{\infty}d\delta~p_{\rm G}(\delta,M)}.
\end{equation}
Another way of quantifying non-Gaussianity is the skewness
\begin{equation}
\label{eq:skew0}
\mu_{3}(M)=\langle \delta^3 \rangle=\int d\delta~p(\delta,M) \delta^3
\end{equation}
To quantify non-Gaussianity from CMB measurements, several authors
\citep[eg.][]{2003ApJS..148..119K,2001MNRAS.325..412V} have considered
the non-linear coupling $f_{NL}$ of the primordial
(pre-recombination) potential field $\Phi$ to the Gaussian field
$\Phi_G$, defined as
\begin{equation}
\label{eq:f_nl}
\Phi=\Phi_G+f_{NL}[\Phi_G^2-\langle\Phi^2\rangle].
\end{equation}
This parametrisation is motivated by a class of inflation models.  A
conversion between $f_{NL}$ and the skewness $\mu_{3}$ of the
(post-combination) density field can be derived using the results of
\cite{2000ApJ...541...10M} and \cite{2001MNRAS.325..412V}.  Current
constraints from galaxy clusters \citep{2003astro.ph..6293A,2000ApJ...532....1R} yield $\alpha \la 4$ at 95\%CL. As we will see in \S\ref{cmb}, the recent WMAP
CMB measurement \citep{2003ApJS..148..119K,2000ApJ...532....1R} translates into $\alpha
\la 1.6$ at 95\%CL for $M \simeq 10^{14} M_{\odot}$.

As a special case, we consider scale free models, where the shape
of the distribution is independent of the scale R. In this case, the
PDF can be written as
\begin{equation}
\label{eq:pdf_noscale}
p(\delta,M)=\sigma(M)^{-1}g\left(\sigma(M)^{-1}\delta\right),
\end{equation}
where $g$ is a function which obeys
\begin{equation}
\label{eq:pdf_cond_noscale}
\int_0^{\infty} dx~g(x)=1, ~~\int_0^{\infty} dx~xg(x)=0, ~~\int_0^{\infty} dx~x^2g(x)=1,
\end{equation}
 so as to verify
equations~(\ref{eq:pdf_cond1}-\ref{eq:pdf_cond2}). For the Gaussian case,
\begin{equation}
\label{eq:g_gauss}
g_{\rm G}(x)= (2\pi)^{-\frac{1}{2}} e^{-x^2/2}.
\end{equation}

\begin{figure}
\psfig{figure=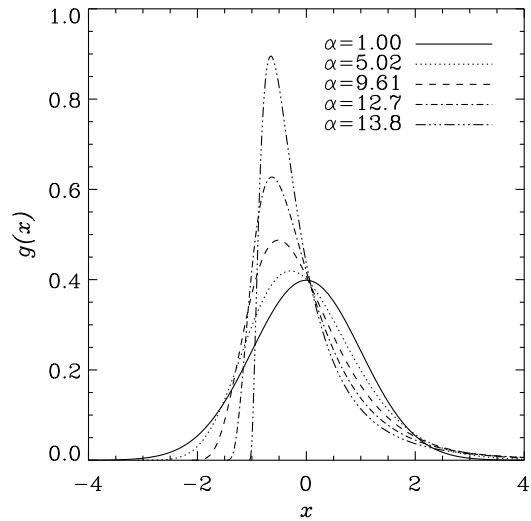,width=80mm}
\caption{Log-normal distribution for several values of the
non-Gaussian parameter $\alpha$. The Gaussian case corresponds to
$\alpha=1$, which is reached as $\beta \rightarrow 0$.  For
$\alpha=5.02$, $\alpha=9.61$, $\alpha=12.7$, and $\alpha=13.8$, the
corresponding $\beta$ values are 0.2, 0.4, 0.6 and 0.8, respectively. For
PDFs with a variance of 1 these correspond to $\mu_3$= 0.61, 1.32, 2.26
and 3.68.}
\label{fig:pdf}
\end{figure}

As an example of a scale invariant distribution, we consider the
log-normal distribution which provides an adequate description of a
range of non-Gaussian models based on non-linear or multi-field
inflation and cosmic strings \citep{2000MNRAS.311..781R}.
For this distribution,
\begin{equation}
\label{eq:lognormal}
g(x)=\frac{1}{\sqrt{2\pi}(x+x_0)
  \beta}\exp\left[{-\frac{\ln^2\left(\frac{x+x_0}{\gamma}\right)}
   {2\beta^{2}}}\right],
\end{equation}
where
\begin{equation}
x_0=\left( e^{\beta^2}-1 \right)^{-\frac{1}{2}},~~
\gamma=\left( e^{2\beta^2}-e^{\beta^2} \right)^{-\frac{1}{2}},
\end{equation}
so as to enforce equation~(\ref{eq:pdf_cond_noscale}). The parameter
$\beta$ controls the non-Gaussianity of the PDF and tends towards zero
as the distribution becomes Gaussian.  Figure \ref{fig:pdf} shows this
distribution for several values of $\beta$ and corresponding values of
$\alpha$.

\section{Generalised Halo Model}
\label{halo_model}
A number of tools exist for modelling the growth of structure since
inflation.  On the largest scales, fluctuations in density are
sufficiently small that their growth can be accurately modelled using
linear theory.  On small scales, however, the complex nonlinear
interactions of gravitational in fall can, at present, only truly be
modelled by using N-body simulation.  These numerical simulations are
however CPU intensive, making the exploration of parameter space
prohibitively slow.  To circumvent this, semi-analytic tools have been
developed to provide fast and accurate fits to numerical simulations.

One such tool is the halo model
\citep{2000ApJ...543..503M,2000MNRAS.318..203S}, which can be used to
calculate the full non-linear power spectrum for a given cosmological
model and has been shown to be in good agreement with high resolution
numerical simulations. The halo model breaks the power spectrum into
two parts: one that dominates on large scales and is determined by the
clustering of halos of different masses, and a second term that is
dominant on small scales and is determined by the density profile of
the halos. The halo model thus requires three pieces of information:
the mass-function of halos, the bias of halos compared to the
background density field and the halo density profile.  The halo model
as described by \cite{2000ApJ...543..503M} and
\cite{2000MNRAS.318..203S} calculates these three ingredients by
assuming that the density perturbations are Gaussian. We show how this
can be extended to include models that have non-Gaussian fluctuations.

\subsection{Mass Function}
\label{massfunction}
\cite{2000MNRAS.311..781R} have proposed a generalisation of the
\cite{1974ApJ...187..425P} formalism which provides a good fit to
N-body simulations for a class of non-Gaussian models. In their
formalism, the fraction of objects of a given mass $M$ or larger that
collapse is calculated by integrating the PDF of the density
fluctuations as
\begin{equation}
\label{eq:f_m}
F(>M)=A\int_{\delta_c}^{\infty}P(\delta,M)d\delta,
\end{equation}
where $A$ is a normalisation factor and $\delta_{c}$ is a critical
over density which can be computed for arbitrary cosmologies using the
results of \cite{1993MNRAS.262..627L} and
\cite{1996MNRAS.282..263E}. The mass function, defined as the number
density of halos per unit mass, is then given by
\begin{equation}
\label{eq:dndm}
\frac{dn}{dM}=\frac{\overline{\rho}}{M}\frac{d\nu}{dM}f(\nu),
\end{equation}
where
\begin{equation}
\label{eq:f_nu}
f(\nu)=\left| \frac{dF}{d\nu} \right|
\end{equation}
and
\begin{equation}
\label{eq:nu}
\nu=\frac{\delta_{c}}{\sigma(M)}.
\end{equation}
The normalisation $A$ is chosen to ensure that all the matter is
accounted for, so that
\begin{equation}
\frac{1}{\langle \rho \rangle} \int_0^{\infty} dM~M\frac{dn}{dM}=
  \int_0^{\infty} d\nu~f(\nu)=1.
\end{equation}

For a scale invariant PDF (Eq.~[\ref{eq:pdf_noscale}]), it is easy to show
that equations~(\ref{eq:f_m}) and (\ref{eq:f_nu}) reduce to
\begin{equation}
f(\nu)= A g(\nu),
\end{equation}
with $A=\left[\int_0^{\infty} dx~g(x)\right]^{-1}$. This simple form
reveals that the mass function at mass M provides a measure of the
normalised PDF $g(x)$ at $x=\nu=\delta_c/\sigma(M)$. This assumes
of course that the generalised Press-Schechter formalism is valid.

For the Gaussian case (Eq.~[\ref{eq:g_gauss}]), we obtain the usual
result
\begin{equation}
\label{eq:f_gauss}
f_{\rm G}(\nu)=\sqrt{\frac{2}{\pi}}e^{-\nu^2/2},
\end{equation}
with the usual normalisation factor $A=2$ prescribed by \cite{1974ApJ...187..425P}.

\begin{figure}
\psfig{figure=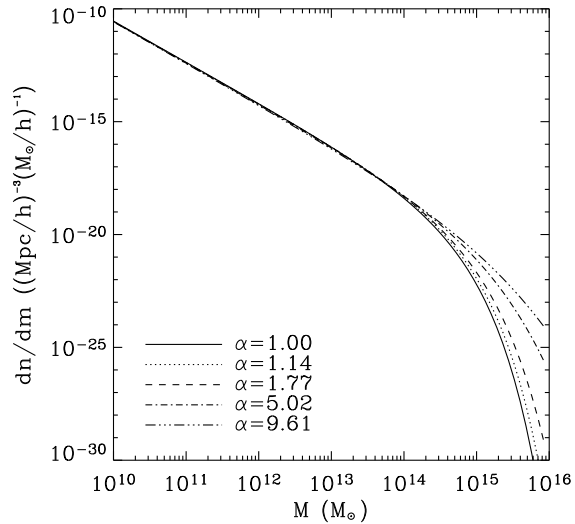,width=80mm}
\caption{Mass function at $z=0$ for the log-normal model with
different values of the non-Gaussian parameter $\alpha$. The Gaussian
case corresponds to $\alpha=1$, and the other values of $\alpha$
correspond to $\beta=0.01$, $0.05$, $0.2$, and $0.4$.  For PDF's with a
variance of 1 these in turn correspond to $\mu_3$=0.03, 0.15, 0.61 and 1.32}
\label{fig:numden_ng}
\end{figure}

For the log-normal distribution (Eq.~[\ref{eq:lognormal}]), we obtain
\begin{equation}
f(\nu)=\frac{A}{\sqrt{2\pi}(\nu+x_0)\beta}
  \exp\left[-\frac{\ln^2(\frac{\nu+x_0}{\gamma})}{2\beta^{2}}\right],
\end{equation}
with
\begin{equation}
A=\left[{\rm erfc}\left(\frac{\ln(x_0/\gamma)}{\sqrt{2}\beta}\right)\right]^{-1},
\end{equation}
where ${\rm erfc}$ is the complementary error function. Figure
\ref{fig:numden_ng} shows the resulting differential mass function
$dn/dM$ for this model at $z=0$, for several values of the
non-Gaussian parameter $\alpha$. Non Gaussianity with $\alpha >1$
tends to increase the number of halos with $M \ga 10^{14} M_{\odot}$
and, to a smaller degree, to deplete halos with masses smaller than
this limit. The high sensitivity of high mass halos to non-Gaussianity
is expected since these halos probe the positive tail of the
distribution. Throughout this paper we assume a standard $\Lambda$CDM
model, with $\Omega_m=0.3,~\Omega_{\Lambda}=0.7,~h=0.7,~n=1 \rm~ and
~\Gamma=0.21$. For this figure we adopted $\sigma_8=0.77$, but we will 
choose other values for this parameter depending on the data set considered.

\subsection{Halo Bias}

The bias parameter in the halo model describes how halos cluster
compared to the underlying linear density field. Generalising the
formalism of \cite{1996MNRAS.282..347M}, \cite{1999MNRAS.310.1111K}
showed that the bias parameter for a general scale invariant PDF is
given by
\begin{equation}\label{eq:bias}
b(\nu)=1-\frac{1}{\delta_{c}}\left[1+\frac{d\ln f(\nu)}{d\ln \nu}\right].
\end{equation}
Note that this expression satisfies the condition $\int
d\nu~f(\nu)b(\nu)=1$, which is required to ensure that the linear
power spectrum is recovered on large scales \citep[see discussion
in][]{2000MNRAS.318..203S}. For the scale invariant case, $f(\nu)$ is
proportional to the PDF (see \S\ref{massfunction}). The halo bias can
therefore be thought of as a measure of the derivative of the PDF in
this case. The measurement of halo bias, using the halo correlation
function for instance, can thus be used to probe the PDF independently
of the mass function.

For the Gaussian case (Eq.~[\ref{eq:f_gauss}]), this reduces to the usual
Mo \& White result
\begin{equation}\label{eq:bias_g}
b_{\rm G}(\nu)=1+ \frac{\nu^{2}-1}{\delta_{c}}.
\end{equation}
More accurate bias functions \citep[eg.][]{1998ApJ...503L...9J} have been devised to provide better fits to N-body
simulations, but they cannot be readily generalised to non-Gaussian
models. Since we are only interested in the effect of
non-Gaussianity relative to the Gaussian case, the above expression
suffices for our purposes.

\begin{figure}
\psfig{figure=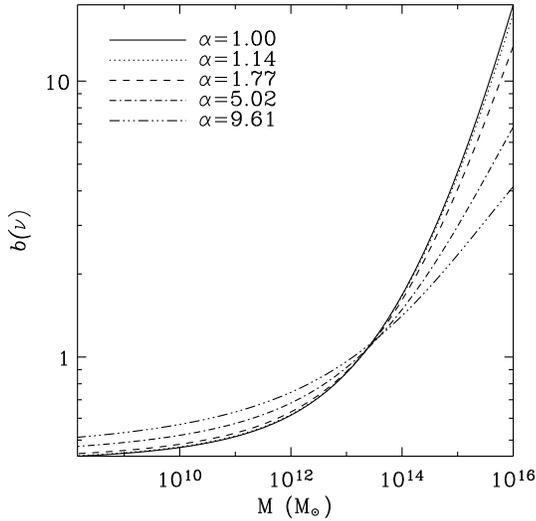,width=80mm}
\caption{Effect of $\alpha$ on the halo bias.  The solid line shows
the Gaussian case.  We see that as we increase the level of
fluctuations in the 3$\sigma$ tail, massive objects (high $\nu$)
become less clustered, while less massive objects (low $\nu$) become
more clustered.}
\label{fig:bias_ng}
\end{figure}

For the log-normal distribution, the bias function is
\begin{equation}
\label{eq:b}
b(\nu)=1+\frac{1}{\delta_{c}} \frac{\nu}{\nu+x_0}
  \left[ \frac{1}{\beta^2}\ln\left(\frac{\nu+x_0}{\gamma}\right)
   -\frac{x_0}{\nu}  \right],
\end{equation}
which does indeed go to Eq. [\ref{eq:bias_g}] as $\beta$ tends towards
zero. Figure \ref{fig:bias_ng} shows the bias as a function of mass
for several values of $\alpha$. For skew-positive PDFs ($\alpha > 1$),
objects with low (high) masses are more (less) clustered compared to the
Gaussian case.

\subsection{Halo profile}
\label{profile}
We model the density profile of the halos as
\begin{equation}
\label{eq:profile}
\rho(r)=\rho_s u(r/r_s),~~ u(x)=x^{-\mu} (1+x)^{\mu-3}
\end{equation}
where $r_s$ is a characteristic radius. The inner slope $\mu$ must be
less than 3 for convergence, and is equal to 1 in the case of a
Navarro, Frenk \& White (\citeyear{1996ApJ...462..563N}, NFW) profile
and 1.4 in the case of a \cite{1998ApJ...499L...5M} profile.  We fix
the normalisation $\rho_s$ so that the density contrast within the
viral radius $r_v$ is $\delta_v=200$, i.e. so that $M=4\pi
\overline{\rho} \delta_v r_v^3/3 =4\pi
\int_0^{r_v}dr~r^2\rho(r)$. This yields $\rho_s=\overline{\rho}
\delta_v c^3 U^{-1}(c)/3$ where we have defined the compactness
parameter $c=r_v/r_s$, and
\begin{equation}
U(c)=\int_0^{c} dx~x^2 u(x)=\frac{c^{3-\mu}}{3-\mu}~_{2}F_{1}(3-\mu,
  3-\mu; 4-\mu; -c)
\end{equation}
where $_{2}F_{1}$ is the hypergeometric function which can be evaluate
numerically \citep{numrecipes}. In the case of the NFW
profile ($\mu=1$), this function reduces to $U(c)=\ln(1+c)-c/(1+c)$.

For the \lcdm Gaussian model, a good fit to the non-linear power
spectrum is given for an NFW profile ($\mu=1$)  with the
concentration parameter is chosen as
\citep{2000PhRvD..62j3506C,2000ApJ...535L...9C}
\begin{equation}
\label{eq:c_mz}
c(M,z)=c_*(z) \left[ \frac{M}{M_{*}(z)} \right]^{-h(z)}
\end{equation}
where $c_*(M,z)\simeq 10.3(1+z)^{-0.3}$ and $h(z) \simeq 0.24 (1+z)^{-0.3}$.
Here $M_*(z)$ is the non-linear mass scale defined by $\nu(M_*,z)=1$.

\cite{2003astro.ph..6293A} have used numerical simulations to measure
the halo profiles in non-Gaussian models. For a model with $\alpha
\approx 3$, they find an increase in the compactness parameter $c$ of
about 20\% compared to the Gaussian case for halos with masses in the
range $10^{11}~\rm to ~10^{15}~h^{-1}M_\odot$. They also find the
inner slope to be steeper ($\mu \approx 1.6$) compared to the Gaussian
NFW case ($\mu = 1$).

More simulations would be needed to derive new scaling relations for
halos in general non-Gaussian models. For our purposes, we use the use
the NFW profile as a reference, but we also study the effect of changing the
compactness and inner slope to those found by
\cite{2003astro.ph..6293A}.

\subsection{Non-linear Power Spectrum}

\cite{2000ApJ...543..503M} and \cite{2000MNRAS.318..203S} showed that
the halo model provides an accurate prescription to compute the
non-linear power spectrum for Gaussian models. In this formalism, the
non-linear power spectrum $P(k)$ at a given redshift is written as
\begin{equation}
\label{eq:powerspec}
P(k)=P_{1}(k) + P_{2}(k),
\end{equation}
where the 1-halo term is
\begin{equation}\label{eq:1halo}
P_{1}(k)=\int_{0}^{\infty} dM ~\frac{dn}{dM} \left[
  \frac{\tilde{\rho}(k,M)}{\overline{\rho}} \right]^{2}
\end{equation}
and the 2-halo term is
\begin{equation}\label{eq:2halo}
P_{2}(k)= \left[ \int_{0}^{\infty} dM ~\frac{dn}{dM} b(M)
  \frac{\tilde{\rho}(k,M)}{\overline{\rho}} \right]^{2} P_{\rm lin}(k).
\end{equation}
In these expressions, $\tilde{\rho}(k,M)$ is the radial Fourier
transform of the density profile (Eq.~[\ref{eq:profile}]) of a halo of
mass $M$. The details of our implementation of the halo model can be
found in \cite{2002PhRvD..66d3002R}.

\begin{figure}
\psfig{figure=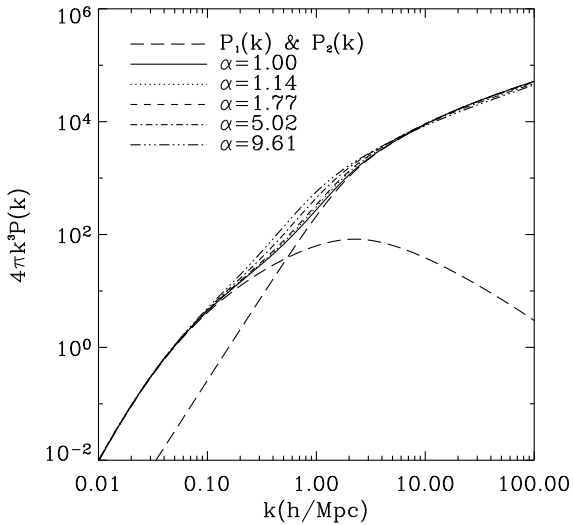,width=80mm}
\caption{Non-linear power spectrum at $z=0$ for several values of $\alpha$.
The $\alpha=1$ Gaussian model (solid line) is a standard $\Lambda$CDM model
with $\Omega_m=0.3$, $\Gamma=0.21$ and $\sigma_8=1.0$. Its decomposition
into the 1-halo and the 2-halo terms are shown as long dashed lines.}
\label{fig:powerspec_ng}
\end{figure}

\begin{figure}
\psfig{figure=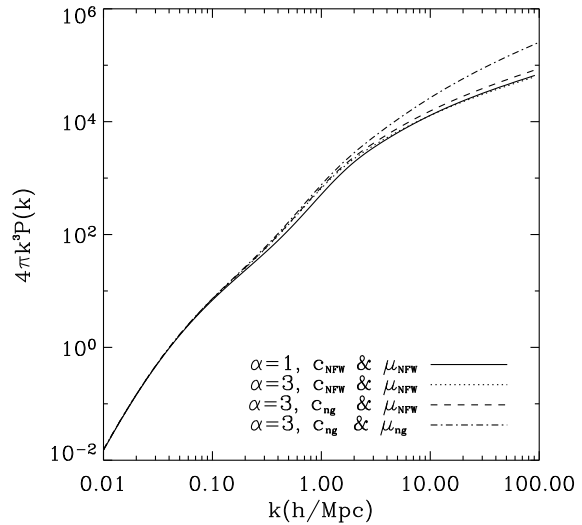,width=80mm}
\caption{Effect of the halo profile on the non-linear power spectrum.
For $\alpha=3$,  Avila-Reese et al. (2003) find an increase in the
compactness parameter, $c_{ng}$ that is roughly 20$\%$ greater than
that for the NFW profile, $c_{NFW}$. They also find the inner slope
$\mu_{ng} \approx 1.6$
to be more cuspy for this level of non-Gaussianity, than that for the
standard NFW profile $\mu_{NFW}=1$.}
\label{fig:power_prof_ng}
\end{figure}

Figure~\ref{fig:powerspec_ng} shows the effect of non-Gaussian initial
conditions on the non-linear power spectrum. Non-Gaussianity is seen
to affect transition scales $0.1\la k \la 3~h$ Mpc$^{-1}$ between the
linear and the non-linear regime. The linear ($k\la 0.1 h$ Mpc$^{-1}$)
and strongly non-linear ($k\ga 3 h$ Mpc$^{-1}$) are not significantly
affected. Figure~\ref{fig:power_prof_ng} shows the effect of varying
the inner slope $\mu$ and the concentration parameter $c$ as indicated
by the non-Gaussian simulations of \cite{2003astro.ph..6293A}. Their
suggested alteration of the concentration parameter by 20\% (which we
applied to all masses) tends to only slightly increase the power
spectrum on small scales ($k\ga 3 h$ Mpc$^{-1}$). On the other hand,
the alteration of the inner profile $\mu$ by 60\% increases the power
spectrum significantly on these scales. Note however, that the mass
dependence of $c(M)$ which we used as a reference
(Eq.~[\ref{eq:c_mz}]) was chosen to reproduce the non-linear power
spectrum in numerical simulations for an NFW profile ($\mu=1$). A 
selfconsistent fit to the non-linear power spectrum from non-Gaussian
simulations would be needed to establish whether this effect is indeed
so pronounced. In the following, we keep $c$ and $\mu$ unchanged from
the Gaussian NFW case, but remain cautious in the interpretation of
our model for the power spectrum on small scales.

\section{Power Spectrum Normalisation}
\label{observations}

\subsection{Cosmic Microwave Background}
\label{cmb}
Using the recent measurements from the WMAP mission,
\cite{2003ApJS..148..175S} found $\sigma_8=0.9 \pm 0.1$ (68\%CL). When
combined with other CMB missions, galaxy surveys and Lyman alpha
surveys, the authors find $\sigma_8=0.84 \pm 0.04$
(68\%CL). The CMB power spectrum provides a 2-point
statistics of the matter fluctuations in the linear regime and is
thus not affected by non-Gaussianity\footnote{Strictly speaking,
non-Gaussianity may slightly bias the algorithms used to measure the
CMB power spectrum, which often assume exact Gaussian statistics. In
this paper, we ignore this small effect.} This is also true when it is
combined with measurement of the galaxy power spectrum on linear
scale, while assuming that the galaxy bias is scale independent.  As a
result, the determination of $\sigma_8$ based on fits to the CMB power
spectrum (with and without galaxy surveys) is insensitive to
non-Gaussianity.

The measurement of higher order statistics of CMB maps, however, does
currently present tight constraints on the level of primordial
non-Gaussianity.  Using the WMAP measurements,
\cite{2003ApJS..148..119K} were able to place limits on the
non-Gaussian coupling defined in equation~(\ref{eq:f_nl}) of
$-58<f_{NL}<134$ (95\%CL). According to
\citet[~figure~2]{2000ApJ...541...10M}, the upper limit corresponds
to an upper limit on the skewness of the post-recombination density
field (Eq.~[\ref{eq:skew0}]) on scale of $M=10^{14}~M_{\odot}$ of
$\mu_{3} \la 8.0\times10^{-2}$, for a \lcdm model with
$\sigma_8=1$. For the log-normal distribution
(Eq.~[\ref{eq:lognormal}]), this corresponds to $\alpha \la 1.6$ at
the 95\% confidence level. Note that this conversion assumes that the
scale dependence of the non-Gaussianity is that of models given by
equation~(\ref{eq:f_nl}).

\subsection{X-ray Cluster Counts}
Counting X-ray clusters is a direct probe of the mass function and
provides a sensitive measure of $\sigma_8$. The main difficulty is to
estimate the masses of X-ray selected clusters.  The most reliable
method to do so consists of measuring the X-ray temperature of the
cluster and to use a mass-temperature ($M$-$T$) scaling relation.
Using an M-T relation derived form hydrodynamical simulations,
several groups have derived a value of $\sigma_8$ around 0.9 from
measurements of the X-ray temperature function \cite[see][~and
reference therein]{2001MNRAS.325...77P}. More recently, the use of
an observational $M$-$T$ relation led to a revision of $\sigma_8$ in
the range $0.7$-$0.8$ \cite[see
eg.][]{2003MNRAS.342..163P,2001astro.ph.11362S}. The uncertainty in
the $M$-$T$ relation is, at present, the largest source of
uncertainty in cosmological parameter estimates using this
method.

\begin{figure}
\psfig{figure=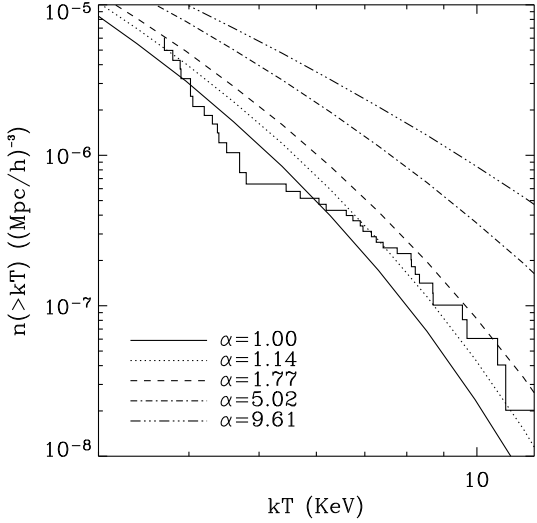,width=80mm}
\caption{Observed temperature function from the compilation of
  Pierpaoli et al. (2001). Also shown is the predicted
  temperature function for several values of the non-Gaussian
  parameter $\alpha$ and for $\sigma_8=0.77$.}
\label{fig:numden_temp}
\end{figure}

In order to study how the determination of $\sigma_8$ using this
method is affected by non-Gaussianity, we consider the compilation
of cluster temperature measurements presented by
\cite{2001MNRAS.325...77P}. The catalogue consists of temperature
measurements for 38 clusters taken using ROSAT and ASCA.  The
cluster temperatures range from $kT=3.5$ to $12$ keV, with those
with temperatures greater than $7.5$ keV forming a volume limited
sample. The resulting cumulative X-ray temperature function $n(>kT)$
from this compilation is shown in figure \ref{fig:numden_temp}. To
compute the cluster masses, \cite{2001MNRAS.325...77P} have used the
$M$-$T$ relation,
\begin{equation}\label{eq:mtrel}
\frac{M(T,z)}{10^{15}h^{-1}M_\odot}=
  \left(\frac{T}{T_*}\right)^{\frac{3}{2}}
  \left(\Delta_{c}E^{2}\right)^{-\frac{1}{2}}
  \left(1-\frac{2\Omega_{\Lambda}}{\Delta_c}\right)^{-\frac{3}{2}},
\end{equation}
where $T$ is the temperature in keV, and $E^2=\Omega_m(1+z)^3+
\Omega_\Lambda+\Omega_k(1+z)^2$. The parameter $\Delta_c$ is the mean
over density inside a virial radius and can be computed using the
fitting functions given by \cite{2001MNRAS.325...77P}. In a later
work, \cite{2003MNRAS.342..163P} fit the observed temperature function
using this relation and find $\sigma_8=0.77_{-0.04}^{+0.05}$ (68\%CL).

\begin{figure}
\psfig{figure=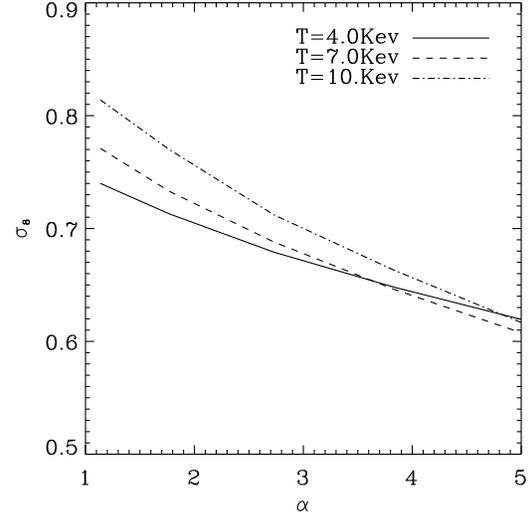,width=80mm}
\caption{Determination of $\sigma_{8}$ from a fit to the
cluster temperature function as a function of the non-Gaussian
parameter $\alpha$. Several temperatures $kT$ for the X-ray cluster
counts are displayed.}
\label{fig:numdenfit}
\end{figure}

Figure~\ref{fig:numden_temp} also shows the predicted temperature
function for different values of the non-Gaussian parameter
$\alpha$. For the masses corresponding to $kT \ga 1$ keV
(corresponding to $M\ga3\times10^{13} h^{-1} M_{\odot}$), a
skew-positive PDF ($\alpha>1$) enhances the cluster counts (see
Figure~\ref{fig:numden_ng}), the effect being more pronounced for
larger temperatures. As a result, the determination of $\sigma_8$ from
the X-ray cluster counts will be affected by
non-Gaussianity. Figure~\ref{fig:numdenfit} shows the value of
$\sigma_8$ derived from a fit to the temperature function as a
function of $\alpha$. Following \cite{2001astro.ph.11362S}, we
considered a fit to $n(>kT)$ at a single temperature $kT$, which we
choose to be 4, 7, and 10 keV, in turn. For $kT \simeq 7$ keV and for
the Gaussian case ($\alpha=1$), we find $\sigma_8\simeq0.77$ in
agreement with \cite{2003MNRAS.342..163P} and
\cite{2001astro.ph.11362S}. The value of $\sigma_8$ required to
fit the temperature function decreases as $\alpha$ increases. The
effect is more pronounced for larger cluster temperatures, as
expected.

\subsection{Cosmic Shear}
\label{shear}
Weak lensing by large scale structure, or cosmic shear, provides a
direct measure of the mass distribution in the universe \cite[see][for
recent
reviews]{2001PhR...340..291B,2003astro.ph..7212R,2003astro.ph..5089V}. This
effect is based on the shear that intervening large-scale structures
induce on the images of background galaxies.  Several groups have now
performed accurate measurements of cosmic shear
\citep{2003AJ....125.1014J,2002astro.ph.10450H,2003MNRAS.341..100B,2002ApJ...577..595H,2002ApJ...572L.131R,2002A&A...393..369V,2002astro.ph..2500V},
which have been used to set constraints on
cosmological parameters.

The 2-point cosmic shear statistics can be described by the weak
lensing power spectrum (see \cite{2000MNRAS.318..625B} for
conventions)
\begin{equation}
C_{\ell}=\frac{9}{16} \left(\frac{H_0}{c} \right)^4 \Omega_m^{2}
\int d\chi ~\left[ \frac{h(\chi)}{ar(\chi)} \right]^2
P\left( \frac{\ell}{r}, \chi \right),
\end{equation}
where $\ell$ is the multipole moment, $\chi$ is the comoving radius,
$r$ is the comoving angular-diameter distance, $a$ is the expansion
parameter, and $H_0$ and $\Omega_m$ are the present value of the
Hubble constant and matter density parameter, respectively. It can be
computed for non-Gaussian models by using the halo-model approximation
for the non-linear power spectrum $P(k,\chi)$ at a given comoving
radius $\chi$, given in equation~(\ref{eq:powerspec}). The radial
weight function is given by
\begin{equation}
h(\chi)=2\int_\chi^{\infty} d\chi'~n(\chi') \frac{r(\chi)r(\chi'-\chi)}
  {r(\chi')},
\end{equation}
where $n(\chi)$ is the red shift distribution of galaxies and is
normalised as $\int d\chi~n(\chi)=1$.

\begin{figure}
\psfig{figure=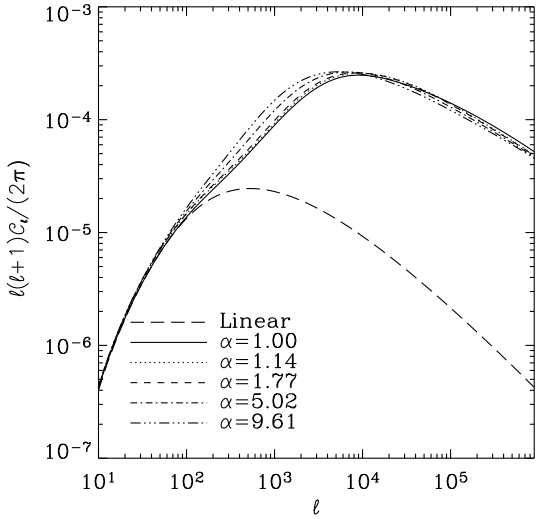,width=80mm}
\caption{Weak lensing power spectrum $C_{\ell}$ for several values
of the non-Gaussian parameter $\alpha$. The galaxy median redshift
was assumed to be $z_m=1$. The Gaussian model ($\alpha=1$) is our
fiducial $\Lambda$CDM model with $\sigma_8=1$.}
\label{fig:cl_ng}
\end{figure}

Figure \ref{fig:cl_ng} shows the weak lensing power spectrum for
several values of $\alpha$. The galaxy redshift distributions was
taken to be $n(z) \propto z^{2} \exp(-(z/z_0)^{1.5})$ with
$z_0=z_m/1.41$ and the median redshift was chosen to $z_m=1$ as
suitable for several cosmic shear surveys. The cosmological parameters
are those for our fiducial model given in \S\ref{massfunction} with
$\sigma_8=1.0$. The linear power spectrum is also plotted for
comparison and reveals that non-linear corrections are very important
on scales $\ell \ga 10^3$. Non-Gaussianity has the most pronounced
effect on intermediate scales $10^2 \la \ell \la 10^4$, which is
consistent with its effect on the 3D power spectrum (see
figure~\ref{fig:powerspec_ng}).  For $\alpha>1$, non-Gaussianity
tends to increase the lensing power spectrum on these scales.

\begin{figure}
\psfig{figure=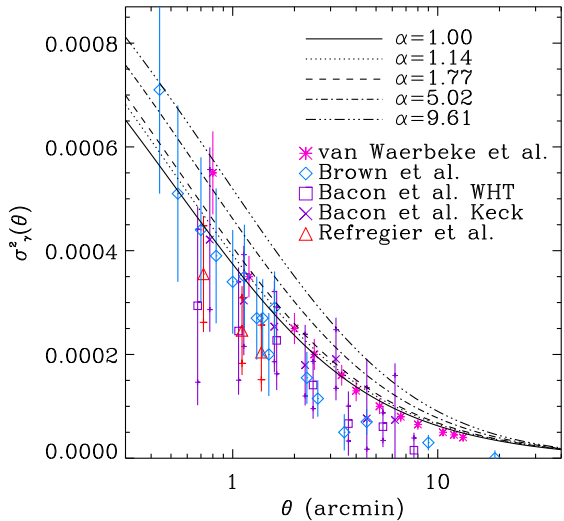,width=80mm}
\caption{Shear variance $\sigma_{\gamma}^2(\theta)$ as a function of
the circular cell radius. Both recent measurements
\citep{2003MNRAS.341..100B,2002ApJ...572L.131R,2002A&A...393..369V,2002astro.ph..2500V}
and the predictions
 for the different non-Gaussian models (with
$z_m=1$ and $\sigma_8=1$) are shown.}
\label{fig:pth_ng}
\end{figure}

In practice, it is often more convenient to consider the shear
variance $\sigma_{\gamma}^2=\langle|\gamma|^2\rangle$ in randomly
place circular cells of radius $\theta$. This is related to the
shear power spectrum by
\begin{equation}
\sigma_{\gamma}^2=\frac{1}{2\pi}\int_0^{\infty} d\ell~C_{\ell} |W_{\ell}|^2
\end{equation}
where $W_{\ell}=2J_{1}(\ell \theta)/(\ell \theta)$ is the Fourier
transform of the cell aperture. The shear variance of the different
non-Gaussian models is shown in figure~\ref{fig:pth_ng}. Also shown
are recent measurements of the shear variance from different surveys
with median galaxy redshifts in the range $0.8 \la z_m \la 1$
\citep{2003MNRAS.341..100B,2002ApJ...572L.131R,2002A&A...393..369V,2002astro.ph..2500V}. The measurements are in rough agreement
with each other and with the Gaussian \lcdm case.  Non-Gaussian
models with $\alpha>1$ produce larger shear variances on scales
$\theta \la 10'$, i.e. on scales to which most surveys are
sensitive. This is not surprising since this limit corresponds to
the transition between the linear ($\theta \ga 10'$) and non-linear
($\theta \la 10'$) regimes for the shear statistics.

\begin{figure}
\psfig{figure=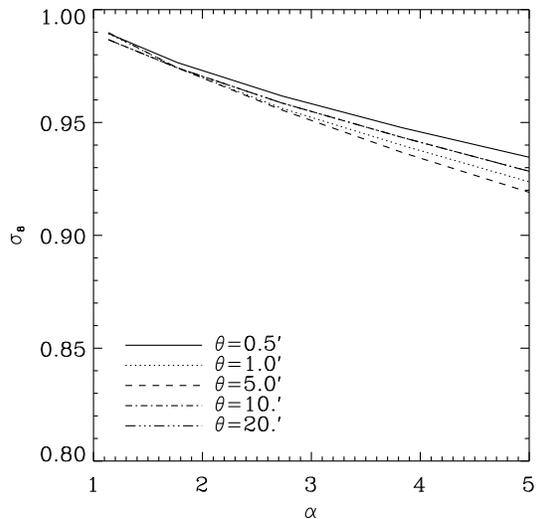,width=80mm}
\caption{Determination of $\sigma_{8}$ from a fit to the
cosmic shear variance $\sigma_{\gamma}^{2}(\theta)$ as a function of
the non-Gaussian parameter $\alpha$. Several scales $\theta$ for the
cosmic shear variance are displayed.}
\label{fig:sigma8_lensing}
\end{figure}

We therefore expect non-Gaussianity to affect the determination of
$\sigma_8$ from current cosmic shear
surveys. Figure~\ref{fig:sigma8_lensing} shows the value of $\sigma_8$
derived from the shear variance measured at different scales and
with $z_m=1$, as a function of the non-Gaussian parameter
$\alpha$. We see that if the universe is non-Gaussian with, say,
$\alpha \approx 3$, the cosmic shear surveys (which assumed
$\alpha$=1) would have overestimated $\sigma_8$ by about 5\%.
This shift is approximately unchanged if the galaxy redshift is
set to $z_m$=0.6, as appropriate for shallower cosmic shear surveys
\citep{2003AJ....125.1014J,2002ApJ...577..595H}, which tend to find
lower values of $\sigma_8$ than deeper ones
(\citealt{2003MNRAS.341..100B,2002ApJ...572L.131R,2002A&A...393..369V,2002astro.ph..2500V};
see however \citealt{2002astro.ph.10450H}). The discrepancy between
shallower and some of the deeper surveys can therefore not be
explained by non-Gaussianity.

As discussed in \S\ref{profile}, the effect of non-Gaussianity
depends on our assumption regarding the halo profile. We find that the
bias in $\sigma_8$ is twice as pronounced if we use the
modified profile suggested by \cite{2003astro.ph..6293A} instead of
the NFW profile. Further high-resolution numerical simulations would
be needed to resolve this theoretical uncertainty.

As expected, the effect of the non-Gaussian parameter, $\alpha$,
depends on the scale being studied.  On large angular scales,
$\theta>10'$, $\alpha$ has a very weak effect on the shear variance, which is
as expected since these scales are still in the linear regime. The
intermediate scales, $1'\la\theta\la10'$, are the most affected by our
non-Gaussian PDF, as can also be seen in figure
\ref{fig:sigma8_lensing}.  This is also consistent with the behaviour
of the weak lensing power spectrum (figure \ref{fig:cl_ng}) since
these intermediate scales correspond to $10^2\la\ell\la 10^3$.  On
scales smaller than $\theta=1'$, the variance from models with
different $\alpha$ values no longer diverge from each other. This
makes measurements on small angular scales less sensitive to
variations in the PDF. On even smaller scales, corresponding to
$\ell\ga10^4$, the shift in the power spectrum changes sign,
suggesting that on these scales the spread in shear variance of our
models will decrease. Since lensing surveys do not probe these
angular scales, we do not investigate this further.

\subsection{Combined Constraints}
As discussed in the previous sections, the determination of $\sigma_8$
from the CMB with and without galaxy surveys is insensitive to
non-Gaussianity. On the other hand, cluster counts and cosmic shear
are affected by non-Gaussianity, as can be seen in
Figures~\ref{fig:numdenfit} and \ref{fig:sigma8_lensing}. For $\alpha
\la 5$, we find that the error in $\sigma_8$ made by ignoring
non-Gaussianity can be approximated by
\begin{equation}
\label{eq:delta_sigma8}
\Delta \sigma_8 = \sigma_8^{G} - \sigma_8 \simeq s (\alpha-1),
\end{equation}
where $\sigma_8^{G}$ is the value derived by (wrongly) assuming a
Gaussian model (i.e. corresponding to $\alpha=1$ in these figures).
The slope $s$ and intercept values $\sigma_8^{G}$ are listed in
table~\ref{tab:delta_sigma8} for cluster counts and cosmic shear at
different temperatures and angular scales. As is apparent from the
figures and in the table, cluster counts are more sensitive ($s$
larger by a factor of about 2) than cosmic shear to
non-Gaussianity. However, as noted above, the use of the
\cite{2003astro.ph..6293A} halo profile instead of the NFW profile
would tend to increase the cosmic shear slope $s$ by about a factor
of 2, which would then nearly match that for cluster abundance.  While
this needs to be established by further numerical simulations, we will
use the NFW slopes $s$ for cosmic shear as a reference but remember
that they are lower limits.

\begin{table}
\begin{center}
\begin{tabular}{|l||r|r|r|}
\hline Method&Scale& $s$ & $\sigma_8^{G}$\\ \hline\hline 
cluster&4.0 keV& 0.032&0.74\\ \cline{2-4} 
counts&7.0 keV& 0.043&0.77\\ \cline{2-4} 
      &10.0 keV& 0.052&0.82\\ \hline\hline 
weak&0.5'& 0.014&0.99\\ \cline{2-4} 
lensing&1.0'& 0.017&0.99\\ \cline{2-4} 
       &5.0'& 0.019&0.99\\ \cline{2-4} 
       &10.0'& 0.015&0.99\\ \cline{2-4} 
       &20.0'& 0.011 &0.99\\
\hline
\end{tabular}
\end{center}
\caption{Fit parameters for the error $\Delta \sigma_8=\sigma_8^{G}
-\sigma_8 \simeq s(\alpha-1)$ made in the measurement of
$\sigma_8$ if non-Gaussianity is ignored.}
\label{tab:delta_sigma8}
\end{table}

If we assume that the PDF shape is scale invariant over all scales, we
have seen (see \S\ref{cmb}) that the recent WMAP skewness measurement
imposes a constraint of $\alpha \la 1.3$ (68\%CL). This limit
corresponds to errors in $\sigma_8$ of $\Delta \sigma_8 \simeq~0.02$
and 0.01, for cluster counts and cosmic shear respectively. These are
significantly smaller to both the current uncertainties
($\Delta\sigma_8 \approx~$0.05 and 0.1, respectively) and the
discrepancy between the different methods and measurements ($\Delta
\sigma_8 \approx 0.3$). Thus, with this assumption, non-Gaussianity can
not explain the internal and external discrepancies between these
methods.

Note however that the above conclusion rests on an extrapolation from
scales probed by WMAP ($\ell \la 1000$ corresponding to $\ga
25~h^{-1}$
 Mpc comoving at decoupling) down to scales relevant for
clusters and cosmic shear, 0.1 to 5.0 comoving $h^{-1}$Mpc. If we drop
the assumption that the PDF is invariant over this large range of
scales, the WMAP measurement no longer constrains the PDF relevant
for these two latter techniques.  In this case, we can attempt to
reconcile the cluster and cosmic shear normalisations using
non-Gaussianity.  We find that this would require $\beta \simeq -0.4$.
This level of negatively skewed PDF does not have a corresponding
$\alpha$ parameter since the distribution does not have a 3$\sigma$
tail in this case.  Such a negative skewness would be inconsistent
with the observed shape of the temperature function (see
figure~\ref{fig:numden_temp}) and of the shear variance (see
figure~\ref{fig:pth_ng}).

Finally, we look at the level of accuracy future surveys need to reach
in order to improve upon the current constraints on non-Gaussianity.
Assuming scale invariance, we have shown that the current $1\sigma$
constraints set by WMAP correspond to uncertainties of $\Delta\sigma_8
\approx 0.02$ for cluster counts and $\Delta\sigma_8\approx0.01$ for
cosmic shear. At present, cluster count measurements have
$1\sigma$ uncertainties of $\Delta\sigma_8 \approx0.05$ which are
dominated by an uncertainty in the normalisation $T_*$ of the $M$-$T$
relation of about 10\%. This means that if the error on $\sigma_8$
from clusters is reduced by a factor of about 2 then they will be
comparable to that arising from our lack of knowledge of the
statistics of the density field. This will be achieved if the
uncertainty in $T_*$ is reduced to about 4\% in future surveys
\citep[see eg.][]{2003MNRAS.342..163P}. Similarly, current lensing
surveys of about 10 deg$^{2}$ have $1\sigma$ errors of
$\Delta\sigma_8\approx0.06$ \citep[eg.][]{2002A&A...393..369V} and
scale as $\sqrt{A}$ where $A$ is the survey area. Lensing surveys will
thus have to increase in sensitivity by a factor of about 6, and thus have
areas of the order of 400 deg$^{2}$, before their measurement of
$\sigma_8$ becomes sensitive to non-Gaussianity. Note, however, that
it may be possible to probe primordial non-Gaussianity by studying
higher order statistics of cosmic shear.

\section{Conclusions}

\label{conclusion}
We have investigated the impact of non-Gaussian conditions on the
determination of $\sigma_8$ via various observational methods. For
this purpose, we have generalised the halo model to
compute cluster statistics and the non-linear matter power spectrum.
In passing, we noted that, for scale free non-Gaussian models, the
halo mass function and bias provide a measure of the linear density
PDF and of its derivative, respectively. The measurement of both of
these statistics, therefore, provides an independent measure of the PDF
and a test of its scale invariance.

The determination of $\sigma_8$ from the CMB power spectrum and from
galaxy surveys in the linear regime is not sensitive to
non-Gaussianity.  On the other hand, the cluster X-ray temperature
function tends to be enhanced at large temperatures ($T \ga 3$ keV)
for skew-positive ($\alpha > 1$) non-Gaussianity. If the primordial
fluctuations are non-Gaussian and this is ignored, the determination
of $\sigma_8$ from current X-ray cluster surveys would be
overestimated by $\Delta \sigma_8
\simeq 0.04(\alpha-1)$.  This depends weakly on the typical
temperature of the clusters.  Cosmic shear statistics from current
surveys probe the non-linear part of the matter power spectrum and
are thus also sensitive to non-Gaussianity. Using the generalised
halo model, we find that the matter power spectrum is enhanced  by
a factor of 2 on scales $0.1\la k \la 3~h\rm~Mpc^{-1}$ for
skew-positive non-Gaussianity of $\alpha=5$. As a result, the cosmic
shear power spectrum is also enhanced by a factor of 2 on scales
$10^2\la \ell \la 10^4$. The error in $\sigma_8$ from current cosmic
shear surveys if non-Gaussianity is ignored is $\Delta
\sigma_8 \simeq 0.02(\alpha-1)$. This behaviour depends weakly on
the angular scale $\theta$ and on the galaxy redshift $z_m$.  We also
find that the nonlinear power spectrum, and therefore cosmic shear
surveys may be sensitive to the halo profile.  This, in turn, may be
sensitive to the primordial density field and needs to be studied in
more detail via high-resolution N-body simulations
\cite[see][]{2003astro.ph..6293A}.

The strongest upper limits on primordial non-Gaussianity are provided
by the recent WMAP measurements and correspond to $\alpha-1 \la
 0.6$
(95\% CL). Within this limit and assuming scale invariance,
non-Gaussianity cannot explain the differences between the $\sigma_8$
values derived from different cosmic shear surveys and X-ray cluster
catalogs. Dropping the assumption of scale invariance from the
scales probed by WMAP to those probed by cluster counts and shear,
non-Gaussianity would tend to decrease the derived values of
$\sigma_8$ for both cosmic shear and cluster counts and thus would
need to be very pronounced to explain the discrepancy. Specifically, a
highly negatively skewed PDF with $\beta\approx-0.4$ (which does not
have a corresponding $\alpha$) would be needed to resolve the
discrepancy, but would not be compatible with the observed shape of
the cluster temperature function and of the cosmic shear 2-point
function. A comparison between future X-ray surveys with a two fold
improvement in cluster mass calibration and future cosmic shear
surveys with 400 deg$^{2}$ will be required to constrain
non-Gaussianity on small scales with a precision matching that of the
current CMB constraints on larger scales. Our results suggest that
the discrepancies are either due to systematics in one or several of
the methods or to non-standard physics other than non-Gaussianity.

\section*{Acknowledgements}
The authors are grateful to Jerry Ostriker, Tarun Saini, Jochen
Weller, Sarah Bridle and Romain Teyssier for useful discussions. AA
was supported by a PPARC student grant and AR was supported in Cambridge
by an Advanced PPARC Fellowship and a Wolfson College Fellowship.

\bibliographystyle{astron}
\bibliography{mybib}
\bsp
\label{lastpage}
\end{document}